\documentclass{ws-procs9x6}

\usepackage{amsmath,amssymb,graphicx,bm}

\newcommand{\im}{\text{Im }}

\newcommand{\cond}[1]{\langle #1 \rangle}

\begin{document}

\title{Structure of the $\sigma$ meson and the softening}

\author{T. HYODO$^*$}

\address{Department of Physics, Tokyo Institute of Technology,\\
Meguro, Tokyo 152-8551, Japan\\
$^*$E-mail: hyodo@th.phys.titech.ac.jp}

\author{D. JIDO}

\address{Yukawa Institute for Theoretical Physics, Kyoto University,\\
Kyoto 606--8502, Japan}

\author{T. KUNIHIRO}

\address{Department of Physics, Kyoto University,\\
Kyoto 606--8502, Japan}

\begin{abstract}
We study the structure of the $\sigma$ meson, the lowest-lying resonance of 
the $\pi\pi$ scattering in the scalar-isoscalar channel, through the 
softening phenomena associated with the partial restoration of chiral 
symmetry. We build dynamical chiral models to describe the $\pi\pi$ 
scattering amplitude, in which the $\sigma$ meson is described either as the 
chiral partner of the pion or as the dynamically generated resonance through 
the $\pi\pi$ attraction. It is shown that the internal structure is 
reflected in the softening phenomena; the softening pattern of the 
dynamically generated $\sigma$ meson is qualitatively different from the 
behavior of the chiral partner of the pion. On the other hand, in the 
symmetry restoration limit, the dynamically generated $\sigma$ meson behaves 
similarly to the chiral partner.
\end{abstract}

\keywords{$\sigma$ meson, chiral dynamics, softening, hadronic molecule}

\bodymatter

\section{Introduction}

The study of the structure of hadron resonances is one of the central issues 
in modern hadron spectroscopy. Among others, the structure of the $\sigma$ 
meson has been intensively studied, since the $\sigma$ meson is considered 
to play an important role in various aspects of hadron and nuclear physics.
For instance, since the scalar-isoscalar excitation of QCD vacuum can be 
regarded as the amplitude fluctuation of the chiral order parameter 
$\langle \bar{q}q\rangle$, the nature of the $\sigma$ meson is crucial to 
understand the dynamical chiral symmetry breaking in 
QCD~\cite{Hatsuda:1994pi}. While the existence of the scalar-isoscalar 
resonance in the $\pi\pi$ scattering is established by recent analyses (see 
e.g. Ref.~\refcite{Caprini:2005zr}), its internal structure is still 
controversial. Thus, it is our aim here to investigate the structure of the 
$\sigma$ meson.

There are many proposals for the structure of the $\sigma$ meson based on 
various effective models: collective $\bar{q}q$ 
excitation~\cite{NJL,Hatsuda:1994pi,HK}, tetraquark structure with strong 
diquark correlation~\cite{Jaffe:1976ig}, dynamically generated $\pi\pi$ 
resonance~\cite{Oller:1997ng}, and so on. Conventionally, the validity of 
certain structure has been tested by comparing the model prediction with 
experimental data, for instance, mass spectrum and the decay properties. 

It is also possible to investigate the structure of the resonance from the 
response to the change of the internal/external parameter of the models. For 
instance, the study of the $N_c$ scaling is successful to disentangle the 
$\bar{q}q$ and other structures for meson resonances~\cite{Pelaez:2003dy}. 
Following this philosophy, we would like to study the spectral change of the 
$\sigma$ meson when the chiral symmetry is partially restored, in order to 
discriminate the different internal structures.

In association with chiral symmetry restoration, the softening of the 
$\sigma$ meson has been discussed~\cite{HK,Hatsuda:1994pi,Hatsuda:1999kd}. 
In the linear realization of chiral symmetry, the $\sigma$ meson forms a 
chiral four-vector together with the pion, and hence they are chiral 
partners. The partial restoration of chiral symmetry induces the softening 
of the $\sigma$ spectrum, which results in the enhancement of the $\pi\pi$ 
cross section in the scalar-isoscalar channel near threshold. It was shown 
later that the threshold enhancement in the $I=J=0$ channel takes place also 
in the nonlinear realization of chiral symmetry without the bare $\sigma$ 
field~\cite{Jido:2000bw,Yokokawa:2002pw} where the $\sigma$ meson is 
expressed as a dynamically generated resonance from the attractive $\pi\pi$ 
interaction. Although similar threshold enhancement of the cross section is 
observed in both cases, the mechanism which causes the softening is quite 
different. Based on these observations, we demonstrate that the softening 
phenomena reflects the structure of the $\sigma$ meson, paying attention to 
the nature of the $s$-wave resonance.

When the symmetry is completely restored, the chiral partners emerge as the 
pair of particles with a degenerate mass. If we regard the mass degeneracy 
as the condition for the chiral partner, we can extend the notion of the 
chiral partner for the dynamically generated $\sigma$ meson in the nonlinear 
realization. We will study the structure of the $\pi\pi$ scattering 
amplitude of our model in the restoration limit, to discuss the chiral 
partner of the pion.

\section{Formulation}

Here we describe the $\pi\pi$ scattering amplitude with the $\sigma$ 
resonance in the $I=J=0$ channel. We consider the low energy behavior of the 
amplitude based on chiral effective Lagrangian, and then introduce the 
unitarity condition to extend the applicability of the model to the 
resonance energy region.

We start from the Lagrangian of two-flavor linear sigma model to derive the 
$\pi\pi$ scattering amplitudes:
\begin{align}
    \mathcal{L}
    =& \frac{1}{4}\text{Tr}
    \left[
    \partial M \partial M^{\dag} -\mu^2 MM^{\dag}
    -\frac{2\lambda}{4!}(MM^{\dag})^{2}
    +h(M+M^{\dag})
    \right] ,
    \label{eq:Lagrangian} 
\end{align}
where $M = \sigma + i\bm{\tau}\cdot \bm{\pi} $. In this Lagrangian, the 
$\sigma$ meson is treated as the chiral partner of the pion. For negative 
$\mu^2$, chiral symmetry is spontaneously broken and three parameters in the 
Lagrangian $\mu$, $\lambda$, and $h$ are related to the chiral condensate 
(pion decay constant) $\cond{\sigma}=f_{\pi}$, the mass of the pion 
$m_{\pi}$, and the mass of the $\sigma$ meson $m_{\sigma}$ in the mean-field 
level. 

Crossing symmetry enables us to express the general $\pi\pi$ scattering 
amplitude as $T_{\text{tree}}(s,t,u)=A(s,t,u)\delta_{ab}\delta_{cd}
+A(t,s,u)\delta_{ac}\delta_{bd}+A(u,t,s)\delta_{ad}\delta_{bc}$, where the 
invariant amplitude $A$ is given, from the Lagrangian~\eqref{eq:Lagrangian} 
at tree level, by~\cite{Jido:2000bw}
\begin{align*}
    A(s)
    =&
    \frac{s-m_{\pi}^2}{\cond{\sigma}^2}
    -\frac{(s-m_{\pi}^2)^2}{\cond{\sigma}^2}
    \frac{1}{s-m_{\sigma}^{2}} .
\end{align*}
In this expression, the first (second) term can be regarded as the leading 
(higher) order contribution in the chiral perturbation theory. The 
coefficient of the leading order term is fixed by the low energy theorem, 
while the low energy constant for the higher order terms is not constrained 
by the symmetry and should be determined by experiments. Thus, we introduce 
a parameter $x$ to express the general amplitude
\begin{align}
    A(s;x)
    =&
    \frac{s-m_{\pi}^2}{\cond{\sigma}^2}
    -x\frac{(s-m_{\pi}^2)^2}{\cond{\sigma}^2}
    \frac{1}{s-m_{\sigma}^{2}} 
    \label{eq:Aamp} .
\end{align}
By choosing $x=1$, we recover the result of the original 
Lagrangian~\eqref{eq:Lagrangian}. If we take $x=0$, then we are left with 
the leading order interaction, which also corresponds to the heavy 
$m_{\sigma}$ limit. In this way, we can smoothly connect the original linear 
sigma model ($x=1$) and the leading order chiral perturbation theory without 
the bare $\sigma$ field $(x=0)$. Projecting Eq.~\eqref{eq:Aamp} onto the 
$I=J=0$ channel, we obtain the tree-level amplitude for the $\pi\pi$ 
scattering as
\begin{align}
    T_{\text{tree}}(s;x)
    =& \frac{m_{\sigma}^{2}-m_{\pi}^2}{\cond{\sigma}^2}
    \Biggl[
    \frac{2s-m_{\pi}^2}{m_{\sigma}^{2}-m_{\pi}^2}(1-x)-5x 
    \nonumber \\
    &-3x\frac{m_{\sigma}^{2}-m_{\pi}^2}{s-m_{\sigma}^{2}}
    -2x\frac{m_{\sigma}^{2}-m_{\pi}^2}{s-4m_{\pi}^2}
    \ln\left(\frac{m_{\sigma}^{2}}{m_{\sigma}^{2}+s-4m_{\pi}^2}\right)
    \Biggr] ,
    \label{eq:Ttree}
\end{align}
in the center-of-mass frame.

Next we consider the unitarity condition $\im T^{-1}(s) = -\Theta(s)/2$ for 
$s>4m_{\pi}^2$, with the two-body phase space function $\Theta(s)=
(16\pi)^{-1}\sqrt{1-4m_{\pi}^2/s}$. Based on the N/D method, we write down 
the general expression of the unitary scattering amplitude $T(s;x)$. 
Matching the chiral interaction $T_{\text{tree}}(s;x)$ with the loop 
expansion of the full amplitude $T(s;x)$, we obtain the amplitude which is 
consistent with both chiral low energy theorem and unitarity 
as~\cite{Oller:1998zr}
\begin{align}
    T(s;x)
    =& \frac{1}{T_{\text{tree}}^{-1}(s;x)+G(s)} 
    \nonumber , \\
    G(s)
    =& \frac{1}{2}\frac{1}{(4\pi)^2}
    \left\{
    a(\mu)+\ln\frac{m_{\pi}^2}{\mu^2}
    +\sqrt{1-\frac{4m_{\pi}^2}{s}}
    \left[\ln
    \frac{\sqrt{1-\frac{4m_{\pi}^2}{s}}+1}
    {\sqrt{1-\frac{4m_{\pi}^2}{s}}-1}
    \right]
    \right\} \nonumber ,
\end{align}
where $a(\mu)$ is the subtraction constant at the subtraction point $\mu$. 
We determine the subtraction constant by excluding the nontrivial CDD pole 
(states which does not originate in the two-body dynamics) in the 
amplitude~\cite{Hyodo:2008xr}
\begin{equation}
    G(s)=0 \quad \text{at} \quad
    s=m_{\pi}^2 
    \label{eq:Gcond} ,
\end{equation}
which leads to $a(m_{\pi})=-\pi/\sqrt{3}$. With this subtraction constant, 
the full scattering amplitude $T$ reduces into the tree level one 
$T_{\text{tree}}$ at $s=m_{\pi}^2$. 

The full amplitude $T(s;x)$ corresponds to the nonperturbative resummation 
of the s-channel loop diagrams up to infinite order. For the $x=1$ case, the 
bare $\sigma$ pole in the Lagrangian acquire the finite width through the 
coupling to the $\pi\pi$ state, and the full scattering amplitude exhibits a 
pole in the complex energy plane. On the other hand, for the $x=0$ case 
without the bare $\sigma$ pole, a resonance can be dynamically generated as 
the pole of the amplitude, if the two-body interaction is sufficiently 
attractive. Indeed, it is shown that the resummation of the leading order 
interaction generates the $\sigma$ meson dynamically~\cite{Oller:1997ti}. In 
the following we compare the properties of these $\sigma$ states: one 
originating in the chiral partner of the pion in the linear $\sigma$ model 
($x=1$), and another generated dynamically from the attractive $\pi\pi$ 
interaction ($x=0$).

\section{Chiral symmetry restoration}

\subsection{Prescription for the symmetry restoration}

Now we introduce the effect of chiral symmetry restoration through the 
modification of the model parameters. It is known that the chiral condensate 
should decrease with the chiral symmetry restoration, so we parametrize the 
condensate by
\begin{equation}
    \cond{\sigma}= \Phi \cond{\sigma}_0 ,
    \nonumber
\end{equation}
where $\cond{\sigma}_0$ is the condensate in vacuum and we vary the 
parameter $\Phi$ from one to zero to express the symmetry restoration. The 
NJL model indicates that $m_{\pi}$ hardly changes as symmetry 
restoration~\cite{Hatsuda:1994pi}, so we assume that it is a constant:
\begin{align}
    m_{\pi}
    = \text{const.} 
    \nonumber
\end{align}
The bare mass of the $\sigma$ should be degenerated with pion when the 
symmetry is restored. This can be achieved by the mean-field relation of Eq.~\eqref{eq:Lagrangian}
\begin{align}
    m_{\sigma}
    =\sqrt{\lambda \frac{\cond{\sigma}^2}{3}+m_{\pi}^2} ,
    \nonumber
\end{align}
with $\lambda$ and $m_{\pi}$ being fixed. 

\subsection{Behavior of the amplitude in the restoration limit}\label{subsec:limit}

It is instructive to study the behavior of the $\pi\pi$ scattering amplitude 
in the limit $\cond{\sigma}\to 0$. 
Here we analytically derive the pole of the amplitude in the restoration limit. In subsection 4.2, we will numerically demonstrate that the obtained result corresponds to the asymptotic behavior of the $\sigma$ pole for $\cond{\sigma}\to 0$.

We first consider the $x=1$ case, where the $\sigma$ meson is the chiral 
partner of the pion. To make the $\cond{\sigma}$ dependence in $m_{\sigma}$ 
explicit, we rewrite the tree level amplitude as
\begin{align*}
    T_{\text{tree}}(s;1)
    =&
    -\frac{5\lambda}{3}
    -\frac{\lambda^2\cond{\sigma}^2}{3}
    \frac{1}{s-m_{\pi}^2-\frac{\lambda}{3}\cond{\sigma}^2} \\
    &-\frac{2\lambda^2\cond{\sigma}^2}{9}
    \frac{1}{s-4m_{\pi}^2}
    \ln \frac{m_{\pi}^2+\frac{\lambda}{3}\cond{\sigma}^2}
    {s-3m_{\pi}^2+\frac{\lambda}{3}\cond{\sigma}^2}
    \label{eq:LSMamp} .
\end{align*}
The second term represents the bare pole of the $\sigma$ meson. As 
$\cond{\sigma}\to 0$, the pole mass of this term decreases and finally it 
coincides with the pion mass. Because of the renormalization 
condition~\eqref{eq:Gcond}, the tree-level amplitude $T_{\text{tree}}(s;1)$
coincides with the full amplitude $T(s;1)$ at $s=m_{\pi}^2$, so the full 
amplitude $T(s;1)$ also has a pole as in the same way with 
$T_{\text{tree}}(s;1)$. Approximating the amplitude by the Breit-Wigner form 
around the pole,
\begin{equation}
    T(s;1)\sim -\frac{g^2}{s-M^2_{\text{pole}}} ,
    \nonumber
\end{equation}
we extract the mass of the state $M_{\text{pole}}$ and the coupling to the 
scattering state $g$. In the present case, we find
\begin{equation}
    g\to 0 ,\quad M_{\text{pole}}\to m_{\pi}
    \quad \text{for}\quad \cond{\sigma}\to 0
    \nonumber .
\end{equation}
This is what we anticipate for the properties of the chiral partner; the 
mass degeneracy with the pion and the vanishing of the coupling constant to 
the $\pi\pi$ state.

Next we consider the $x=0$ case without the bare $\sigma$ pole. In this 
case, $\cond{\sigma}$ dependence of the tree-level 
amplitude~\eqref{eq:Ttree} exclusively stems from the overall factor,
\begin{equation}
    T_{\text{tree}}(s;x)
    \propto \frac{1}{\cond{\sigma}} .
    \nonumber
\end{equation}
Taking the restoration limit $\cond{\sigma}\to 0$, this term diverges, and 
therefore the full amplitude is solely determined by the loop function 
$G(s)$:
\begin{equation}
    T(s;x)=\frac{1}{T_{\text{tree}}^{-1}(s,x)+G(s)}
    \to \frac{1}{G(s)}\quad \text{for}\quad \cond{\sigma}\to 0 .
    \nonumber
\end{equation}
Thus, the pole of the amplitude in the restoration limit is given by the 
zero of $G(s)$. The present renormalization scheme requires $G(s)=0$ for 
$s=m_{\pi}^2$, which indicates the existence of a pole at $\sqrt{s}=m_{\pi}$ 
in the $\sigma$ channel. The coupling constant $g$ can be obtained by 
calculating the residue of the pole:
\begin{align}
    g^2|_{\cond{\sigma}\to 0}
    =&(4\pi)^2
    \left(\frac{\pi}{3\sqrt{3}}-\frac{1}{2}\right)^{-1}m_{\pi}^2 .
    \label{eq:NLScoupling}
\end{align}
Thus, for the dynamically generated $\sigma$ meson, the amplitude has a pole 
at the pion mass with the coupling constant which is proportional to 
$m_{\pi}$ for $\cond{\sigma}\to 0$.

This result has an interesting implication for the chiral partner. Strictly 
speaking, the notion of the ``chiral partner'' is defined only in the chiral 
limit ($m_{\pi}\to 0$), where the SU(2)$\times$SU(2) symmetry is exact in 
the Wigner phase. In the chiral limit, Eq.~\eqref{eq:NLScoupling} indicates 
$g^2|_{\cond{\sigma}\to 0}=0$, so the asymptotic value of the mass and 
coupling constant of the dynamically generated $\sigma$ meson is exactly the 
same with the chiral partner case. Namely, the dynamically generated 
$\sigma$ meson behaves as if it is the chiral partner of the pion, for 
$m_{\pi}\to 0$. 

\section{Numerical study for the softening phenomena}

\subsection{Structure of the $\sigma$ meson in vacuum}\label{subsec:vacuum}

We first show the description of the scattering amplitude in vacuum. We 
choose canonical values of the parameters as $\cond{\sigma}_0=93$ MeV, 
$m_{\pi}=140$ MeV, and $m_{\sigma}=550$ MeV. By taking the parameter $x=1$, 
the $\sigma$ meson is described as the chiral partner of the pion (chiral 
$\sigma$), and we refer to this case as ``model A''. Choosing $x=0$, we 
obtain the dynamically generated $\sigma$ meson in the amplitude. This case 
is called as ``model B''. 

To check the agreement with the physical amplitude, we study the properties 
of these models without symmetry restoration. We calculate the scattering 
length $a=\frac{1}{32\pi}T(4m_{\pi}^2)$ (in units of $m_{\pi}^{-1}$) in 
these models. The results are shown in Table~\ref{tbl:model}, together with 
the pole position of the amplitude in the complex energy plane. We find a 
qualitative agreement with the recent analyses of experimental 
data~\cite{Pislak:2003sv,Caprini:2005zr}, $a_{\text{exp}}\sim 0.216$ and 
$z=441-272i$ MeV.

\begin{table}[b]
    \tbl{Properties of the models: value of parameter $x$,
    possible origin of the pole,
    the scattering length $a$ in units of $m_{\pi}^{-1}$, 
    and the pole position of the amplitude in vacuum.
    }
    {\begin{tabular}{l|ccccc}
    \hline
     & $x$ & origin & $a$ [$m_{\pi}^{-1}$] & pole position [MeV]\\
    \hline
    model A & 1 & chiral partner & 0.244 & $423-126 i$ \\
    model B & 0 & dynamically generated & 0.174 & $364-356 i$ \\
    \hline
    \end{tabular}}
    \label{tbl:model}
\end{table}%

\subsection{Softening of the $\sigma$ meson}\label{subsec:softening}

We then study the variation of the scattering amplitude in the $\sigma$ 
channel along with partial restoration of chiral symmetry. We plot the 
reduced cross section $\bar{\sigma}=|T|^2/s$ and the trace of the pole 
position as functions of the total center-of-mass energy $\sqrt{s}$, by 
changing the parameter $\Phi$ from 1 to 0.

\begin{figure}[tbp]
    \centering
    \includegraphics[width=5cm,clip]{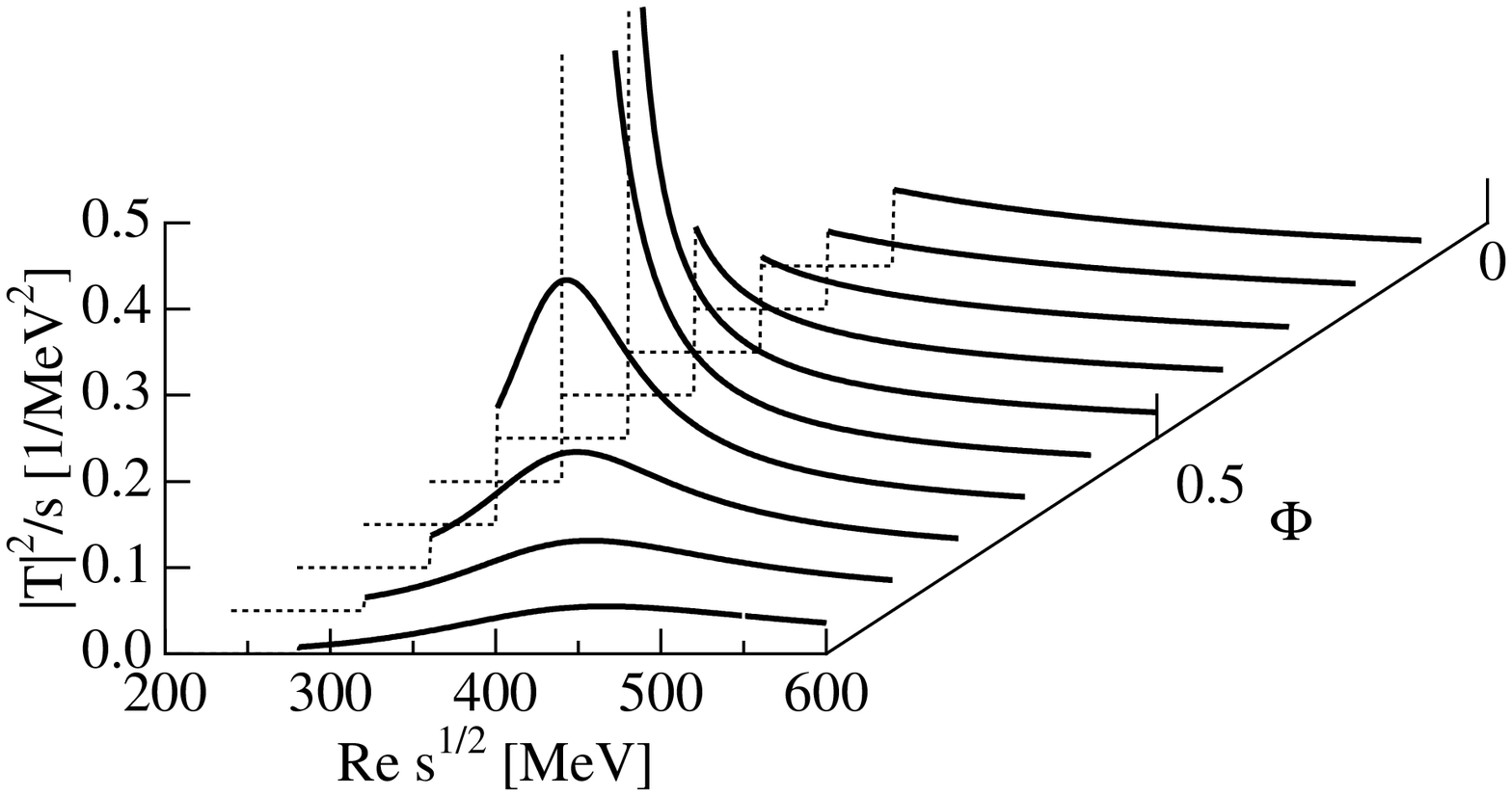}
    \includegraphics[width=5cm,clip]{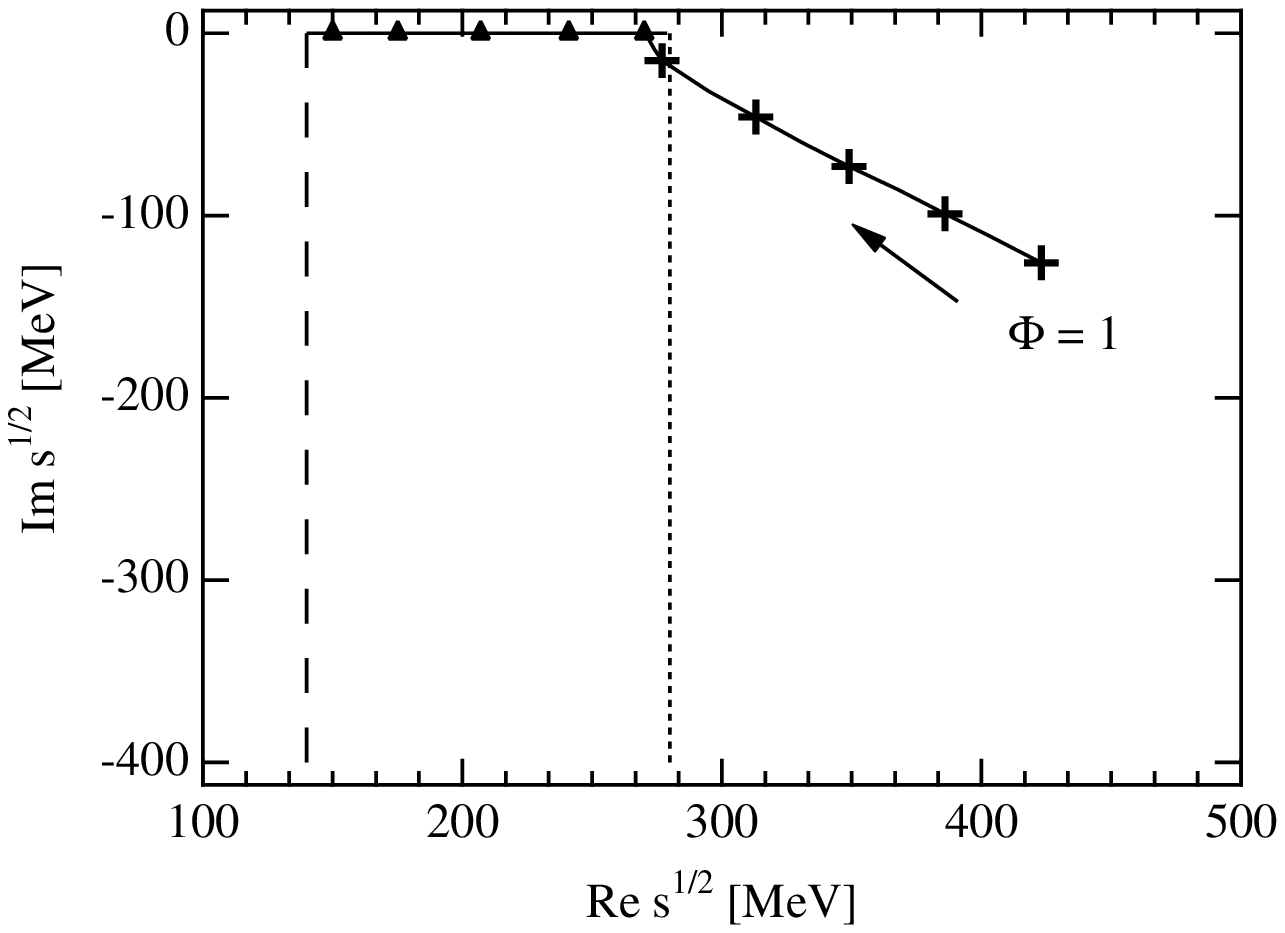}
    \caption{\label{fig:modelA}
    Mass spectra of the $\sigma$ meson (left)
    and the trace of the pole positions (right) in model A ($x=1$). 
    The poles on the first (second) Riemann sheet are plotted by triangles 
    (crosses). Symbols are marked with each 0.1 step of $\Phi$.
    Arrow indicates the direction of the movement of the pole as 
    the parameter $\Phi$ is decreased from 1 to 0.
    Dotted (dashed) line represents the energy corresponds to the threshold 
    (mass of pion).}
\end{figure}

The invariant mass spectra and the pole trajectory of model A are shown in 
Fig.~\ref{fig:modelA}, where the softening of $\sigma$ is clearly observed. 
As the symmetry is restored, the $\sigma$ pole moves toward the $\pi\pi$ 
threshold and the spectrum gets narrow and enhanced around threshold. In 
this case, since the $\sigma$ meson is treated as the chiral partner of 
pion, the softening phenomena is driven by the decrease of the bare mass of 
the $\sigma$ and its consequence of the reduction of the phase space for the 
decay~\cite{Hatsuda:1999kd}. In the limit $\cond{\sigma}\to 0$, the pole 
approaches the mass of the pion, as indicated by the analysis in 
section~\ref{subsec:limit}.

\begin{figure}[tbp]
    \centering
    \includegraphics[width=5cm,clip]{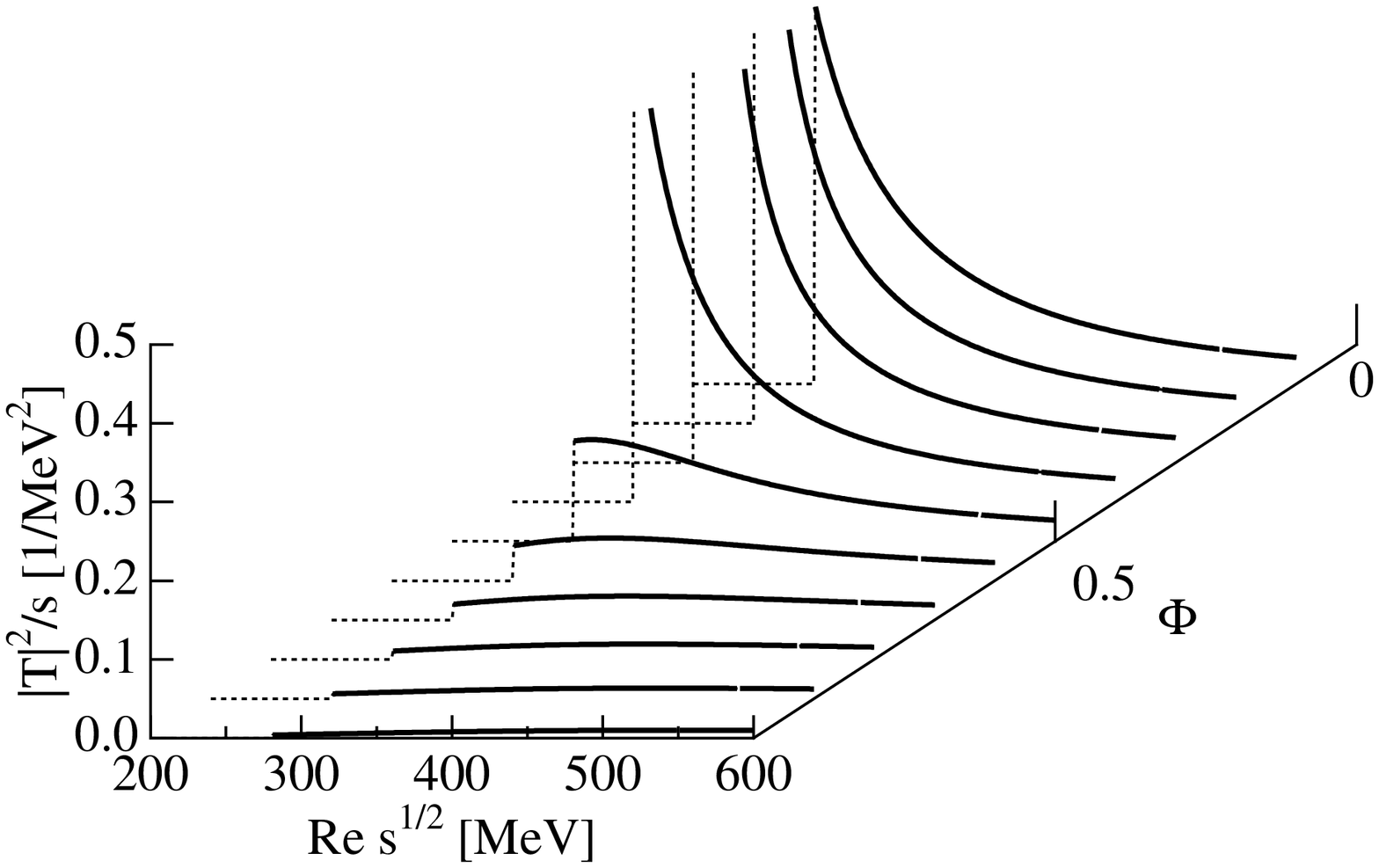}
    \includegraphics[width=5cm,clip]{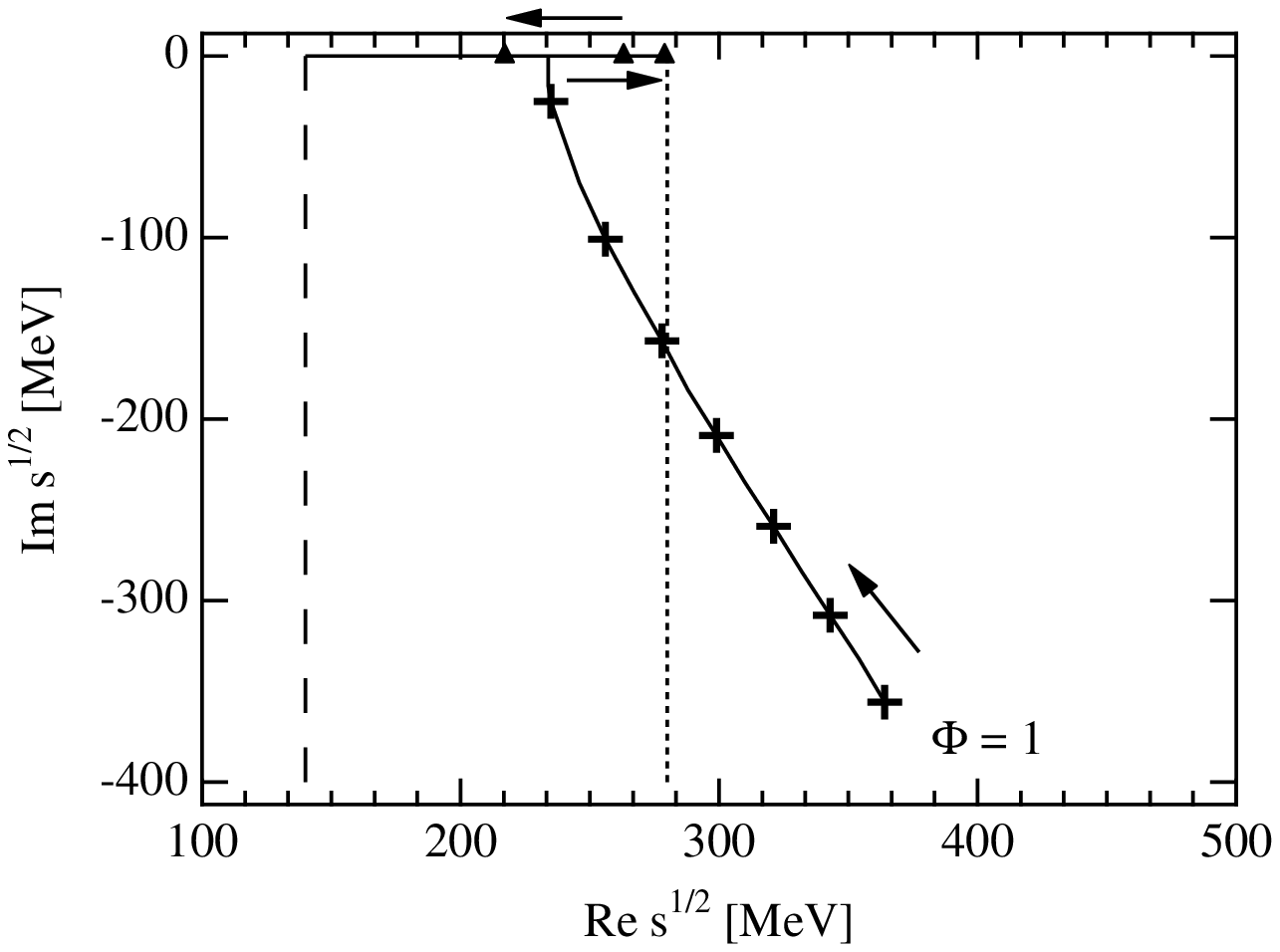}
    \caption{\label{fig:modelB}
    Mass spectra of the $\sigma$ meson (left)
    and the trace of the pole positions (right) in model B ($x=0$). 
    The poles on the first (second) Riemann sheet are plotted by triangles 
    (crosses). Symbols are marked with each 0.1 step of $\Phi$.
    Arrows indicate the direction of the movement of the pole as 
    the parameter $\Phi$ is decreased from 1 to 0.
    Dotted (dashed) line represents the energy corresponds to the threshold 
    (mass of pion).}
\end{figure}

We show the results of model B in Fig.~\ref{fig:modelB}. In this case, 
although the threshold enhancement takes place, the change of the spectrum 
as well as the trace of the pole are qualitatively different from those of 
model A. Especially, the pole in the complex energy plane goes to the 
subthreshold energy region, keeping the width finite. This is a peculiar 
feature of the dynamically generated $\sigma$ meson. As a consequence, the 
strong enhancement of the cross section occurs at much later stage of the 
symmetry restoration, compared with the model A.

Let us discuss how this structure appears in model B. The mechanism of the 
threshold enhancement in the nonlinear $\sigma$ model has been studied in 
Ref.~\refcite{Jido:2000bw}; the partial restoration of chiral symmetry  
induces the reduction of the pion decay constant. Since the low energy 
interaction is proportional to $1/\cond{\sigma}^2$, the symmetry restoration 
effectively increases the attractive interaction. Thus, the dynamically 
generated $\sigma$ resonance will eventually turns into a $\pi\pi$ bound 
state when the interaction becomes sufficiently attractive. At this stage, 
however, it is important to recall that the $\sigma$ is in $s$-wave. In this 
case, when the attraction is increased, the resonance first becomes the 
\textit{virtual state} which is the pole on the second Riemann sheet of the 
energy plane but lies below the threshold~\cite{Bohm:2001}. The peculiar 
pole trajectory in Fig.~\ref{fig:modelB} is due to the appearance of the 
virtual state.

It is worth mentioning that the finite pion mass is important for the 
appearance of the virtual state. If $m_{\pi}= 0$, the $\pi\pi$ threshold 
lies at $\sqrt{s}=0$, so there is no region where the virtual state appears. 
Indeed, the virtual state was not seen in the analysis of the dynamically 
generated $\sigma$ meson in Ref.~\refcite{Yokokawa:2002pw}, studied in the 
chiral limit. In addition, the $s$-wave nature of the $\sigma$ is also 
essential for the virtual state. Therefore, the behavior of the $\rho$ meson 
in $p$-wave amplitude will not exhibit such a structure (see also 
Ref.~\refcite{Hanhart:2008mx}).

As we further restore the symmetry, the $\sigma$ meson becomes the bound 
state. In the limit $\cond{\sigma}\to 0$, the pole moves toward the pion 
mass. This is again in agreement with the result in 
section~\ref{subsec:limit}: the appearance of the pole at the pion mass in 
the restoration limit.

\section{Summary}

We have studied the properties of the $\sigma$ meson in the $\pi\pi$ 
scattering associated with the restoration of chiral symmetry. Two models 
are constructed based on chiral low energy interaction and unitarity of the 
scattering amplitude: one describes the $\sigma$ meson as the chiral partner 
of the pion, and the other treats the $\sigma$ meson as dynamically 
generated resonance. 

For the dynamically generated $\sigma$ meson, we find the qualitative 
difference from the chiral partner $\sigma$ in the softening behavior, 
namely, the movement of the pole of the amplitude and its consequence of the 
change of the spectrum. The difference stems from the mechanism which drives 
the softening, and it is the appearance of the virtual state that leads to 
the distinct behavior of the dynamically generated $\sigma$. 

We also study the asymptotic properties of the $\sigma$ pole in the 
restoration limit. For the $\sigma$ meson as the chiral partner, we find 
that the mass of the $\sigma$ pole approaches the pion mass and the coupling 
to the $\pi\pi$ state vanishes. For the dynamically generated $\sigma$, we 
also find the mass degeneracy with the pion, and the coupling strength 
vanishes in the chiral limit $m_{\pi}\to 0$. Namely, the behavior of the 
dynamically generated $\sigma$ pole in the restoration limit is essentially 
the same with what we expect for the chiral partner. This is a nontrivial 
result which urge us to speculate the possibility of the dynamically 
generated $\sigma$ meson as the chiral partner of the pion.

In this way, through the comparison of two models, we draw two conclusions: 
(i) with partial restoration of chiral symmetry, the difference of the 
internal structures is reflected in the spectral change of the $\sigma$ 
channel, and (ii) in the symmetry restoration limit, the difference of the 
structure is reduced and we obtain essentially the same behavior of the 
$\sigma$ pole for $m_{\pi}\to 0$. More comprehensive analysis, including the 
case with the $\sigma$ meson as the CDD pole, is now 
underway~\cite{Sigmameson}.

\section*{Acknowledgments}

The authors are grateful to M. Oka and Y. Kanada-En'yo for useful 
discussion. This work is partly supported by the Global Center of Excellence 
Program by MEXT, Japan through the Nanoscience and Quantum Physics Project 
of the Tokyo Institute of Technology, the Grant-in-Aid for Scientific 
Research from MEXT and JSPS (Nos. 21840026, 20540273, 2054026, and 
20028004), and the Grant-in-Aid for the Global COE Program ``The Next 
Generation of Physics, Spun from Universality and Emergence'' from MEXT of 
Japan. This work was done in part under the Yukawa International Program for 
Quark-hadron Sciences (YIPQS).


\begin{thebibliography}{10}

\bibitem{Hatsuda:1994pi}
T.~Hatsuda and T.~Kunihiro, {\em Phys. Rept.} {\bf 247}, 221 (1994).

\bibitem{Caprini:2005zr}
I.~Caprini, G.~Colangelo and H.~Leutwyler, {\em Phys. Rev. Lett.} {\bf 96},  132001 (2006).

\bibitem{NJL}
  Y.~Nambu and G.~Jona-Lasinio,
  {\em Phys. Rev.}  {\bf 122}, 345 (1961); {\em Phys. Rev.}  {\bf 124}, 246 (1961).

\bibitem{HK}
  T.~Hatsuda and T.~Kunihiro,
  {\em Prog. Theor. Phys.}  {\bf 74}, 765 (1985);
  {\em Phys. Lett.}  B {\bf 185}, 304 (1987).

\bibitem{Jaffe:1976ig}
R.~L. Jaffe, {\em Phys. Rev.} {\bf D15}, 267 (1977).

\bibitem{Oller:1997ng}
J.~A. Oller, E.~Oset and J.~R. Pelaez, {\em Phys. Rev. Lett.} {\bf 80}, 3452
  (1998).

\bibitem{Pelaez:2003dy}
J.~R. Pelaez, {\em Phys. Rev. Lett.} {\bf 92}, 102001 (2004).

\bibitem{Hatsuda:1999kd}
T.~Hatsuda, T.~Kunihiro and H.~Shimizu, {\em Phys. Rev. Lett.} {\bf 82}, 2840 (1999).

\bibitem{Jido:2000bw}
D.~Jido, T.~Hatsuda and T.~Kunihiro, {\em Phys. Rev.} {\bf D63}, 011901 (2001).

\bibitem{Yokokawa:2002pw}
K.~Yokokawa, T.~Hatsuda, A.~Hayashigaki and T.~Kunihiro, {\em Phys. Rev.} {\bf C66}, 022201 (2002).

\bibitem{Oller:1998zr}
J.~A. Oller and E.~Oset, {\em Phys. Rev.} {\bf D60}, 074023 (1999).

\bibitem{Hyodo:2008xr}
T.~Hyodo, D.~Jido and A.~Hosaka, {\em Phys. Rev.} {\bf C78}, 025203 (2008).

\bibitem{Oller:1997ti}
J.~A. Oller and E.~Oset, {\em Nucl. Phys.} {\bf A620}, 438 (1997).

\bibitem{Pislak:2003sv}
S.~Pislak {\em et~al.}, {\em Phys. Rev.} {\bf D67}, 072004 (2003).

\bibitem{Bohm:2001}
A.~Bohm, {\em Quantum Mechanics: Foundations and Applications} (Springer, New York, 2001).

\bibitem{Hanhart:2008mx}
C.~Hanhart, J.~R. Pelaez and G.~Rios, {\em Phys. Rev. Lett.} {\bf 100}, 152001 (2008).

\bibitem{Sigmameson}
T.~Hyodo, D.~Jido and T.~Kunihiro, arXiv:1007.1718 [hep-ph].

\end{thebibliography}

\end{document}